\documentclass[a4paper,12pt]{article}

\usepackage{graphicx}
\usepackage{amsfonts, amsmath}
\usepackage{amssymb}
\usepackage{mathrsfs}
\usepackage{setspace}
\usepackage[latin1]{inputenc}
\usepackage{comment}
\usepackage{cite}
\usepackage{slashed}
\usepackage{bbold}

\DeclareMathOperator{\Tr}{Tr}

\title{\vspace{-3.5cm}\begin{flushright}
 \small SU-4252-918 \\ IMSc/2011/8/8 \\ Phys. Rev., 2012, D85, 025017 
\end{flushright}
\vspace{3.75cm} Mixed States from Anomalies}

\author{\textbf{A. P. Balachandran}\footnote{bal@phy.syr.edu} \\
\emph{Department of Physics, Syracuse University,} \\
\emph{Syracuse, NY 13244-1130, USA} \\ and \\ \emph{Institute of Mathematical
Sciences,} \\ \emph{CIT Campus, Taramani, Chennai 600113, India} \\ and \\
\emph{International Institute of
Physics (IIP-UFRN)} \\
\emph{Av. Odilon Gomes de Lima 1722, 59078-400 Natal, Brazil}\vspace*{1cm} \\
\textbf{Amilcar R. de Queiroz}\footnote{amilcarq@unb.br} \\ \emph{Instituto de
Fisica, Universidade de
Brasilia,} \\ \emph{Caixa Postal 04455, 70919-970, Brasilia, DF, Brazil}}

\begin{document}

\maketitle

\begin{abstract}
There are several instances where quantum anomalies of continuous and discrete
classical symmetries play an important role in fundamental physics. Examples
come
from chiral anomalies in the Standard Model of fundamental interactions and
gravitational anomalies in string theories. Their generic 
origin is the fact that classical symmetries may not preserve the domains of
quantum
operators like the Hamiltonian. In this work, we show by simple examples that
anomalous symmetries can often be implemented at the expense of working with
mixed
states having non-zero entropies. In particular there is the result on color
breaking by non-abelian
magnetic
monopoles. This anomaly can be rectified by using impure states. We also argue
that
non-abelian groups of twisted bundles are always anomalous for pure states
sharpening
an earlier argument of Sorkin and Balachandran \cite{bal-book-1}. This is the
case of
mapping
class groups of geons \cite{bal-book-1} indicating that \emph{large} diffeos
are anomalous for pure states in the presence of geons. Nevertheless diffeo
invariance may be
restored by using impure states. This work concludes with examples of these
ideas drawn from
molecular
physics.

The above approach using impure states is entirely equivalent to restricting all
states to the algebra of observables invariant under the anomalous symmetries.
For anomalous gauge groups such as color, this would mean that we work with
observables singlet under global gauge transformations. For color, this will
mean that we work with color singlets, a reasonable constraint.
\end{abstract}

\section{Introduction}

There is perhaps a dominant perception that quantum anomalies of classical
symmetries
can occur
only in the context of quantum field theories. Typically they arise in the
course of
regularizing divergent expressions in quantum fields \cite{gaume-1,harvey}, causing the impression
that it is these divergences that cause anomalies.

It is however known that anomalies can occur in simple quantum mechanical
systems
such as a
particle on a circle or a rigid rotor. Esteve \cite{esteve-1, esteve-2}
explained long ago that
the
presence or
otherwise of anomalies is a problem of domains of quantum operators. Thus while
quantum state
vectors span a Hilbert space $\mathcal{H}$, the Hamiltonian $H$ is seldom
defined on
all
vectors of $\mathcal{H}$. For example, the space $\mathcal{H}$ of
square-integrable
functions on $\mathbb{R}^3$ contains non-differentiable functions $\psi$, but
the
Schroedinger
Hamiltonian $H=-\frac{1}{2m}\nabla^2$ is not defined on such $\psi$. Rather $H$
is
defined only on a
dense subspace $D_H$ of $\mathcal{H}$. If a classical symmetry $g$ does
not
preserve
$D_H$,
$gD_H\neq D_H$, then $Hg~\psi$ for $\psi\in D_H$ is an
ill-defined expression.
In
this case,
one says that $g$ is anomalous \cite{esteve-1, esteve-2}. See also \cite{manton,nelson,Balachandran:1996iv,Balachandran:1996jz,Balachandran:1998zq,Aguado:2000aw,ananos}.

In the present work, we explore the possibility of overcoming anomalies by using
mixed
states. There are excellent reasons for trying to do so, there being classical
gauge
symmetries like $SU(3)$ of QCD or \emph{large} diffeomorphisms (diffeos) of
manifolds (see below)
which
can
become anomalous. Color $SU(3)$ does so in the presence of non-abelian
monopoles
\cite{mukunda-1,mukunda-2,mukunda-3}, while ``large'' diffeos do so for suitable
Friedman-Sorkin geon
manifolds
\cite{friedman-witt,friedman-sorkin-1,friedman-sorkin-2}.
It is surely worthwhile to find ways to properly implement these symmetries.

In this paper, we first focus on simple quantum mechanical systems to illustrate
how
the use of
impure states can often restore the anomalous symmetries. We then discuss color
breaking by
non-abelian monopoles. Finally we argue that structure groups of twisted
non-abelian
bundles
are \emph{always} anomalous for pure states. This claim is illustrated with
examples from molecular physics, where such groups are not only compact, but
discrete as well. In later
work, we
will extend these considerations to diffeo anomalies.

While non-abelian structure groups of twisted bundles are always anomalous,
abelian
groups also
of course can be anomalous. For instance, parity anomaly for a particle on a
circle
(discussed in section 2 of this work) and the axial
$U(1)_A$ anomaly in the Standard Model are both abelian. The crucial issue
is
whether
the classical symmetry preserves the domains of appropriate operators like the
Hamiltonian. If
they do not preserve such domains, then they are anomalous. The important
feature of
non-abelian structure
groups of twisted bundles is that they \emph{never} preserve the domain of the
Hamiltonian. More on this later.

Our use of mixed states is entirely equivalent to restricting the algebra of
observables to those invariant under symmetries. For global symmetries, this can
be a restriction, as there may be no good reason to discard non-invariant
observables. But for many \emph{gauge} symmetries, this requirement is often
already implied by gauge invariance. That is the case for mapping class groups
of manifolds and ``symmetries'' of molecules. For the global color group which is emergent from gauge transformations, constraining observables to singlets is
reasonable in view of the hypothesis of color confinement.

In this paper, all examples we work on are those of global anomalies. As a matter of specificity, most of these examples are of ``global'' gauge anomalies like the global color group or ``large'' diffeos.

We shall see that even though we can overcome the problem of implementing a
symmetry, time evolution still does involve the choice of a domain. In this
sense, the theory carries the memory of the anomaly.

But when the anomaly is for a classical symmetry, a domain and its transform by
this symmetry  are \emph{equivalent}, exactly as in the case of
standard spontaneous symmetry breaking. In quantum field theory, there seems to be an
associated Nambu-Goldstone theorem as well. But now we can show that all this can
happen on a spatial manifold with a boundary, and does not require its infinite
volume. We will elaborate on these issues elsewhere.

The present paper is organized as follows: in section 2, we discuss parity and
time
reversal
for a particle on a circle, this being a very simple example; In section 3, we
adapt
this
discussion to color breaking; In section 4, we show the generic nature of our
results.
We finally conclude with examples from molecular physics.

\section{Anomalous Parity and Time Reversal for Particle on a Circle}

\subsection{Classical Theory}

A point on a circle $S^1$ can be described by $e^{i\varphi}$, with $\varphi$
being
real. Its classical equation of motion assuming it to be free is
\begin{equation}
\label{eq:cl-e.o.m-1}
      \frac{d^2}{dt^2}\varphi(t)=0,
\end{equation}
where $t$ labels time.

If $S^1$ is embedded in $\mathbb{R}^2$,
\begin{equation}
      S^1=\{ x=(x_1,x_2)\in\mathbb{R}^2: x_1^2+x_2^2=1\},
\end{equation}
then we can relate $e^{i\varphi}$ to $x$ by writing
\begin{equation}
      x_1+ix_2=e^{i\varphi}.
\end{equation}

The parity transformation $P:(x_1,x_2)\mapsto (x_1,-x_2)$ takes $e^{i\varphi}$
to
$e^{-i\varphi}$, that is,
\begin{equation}
\label{eq:parity-1}
P:e^{i\varphi}\mapsto e^{-i\varphi}.
\end{equation}
It is an orientation-reversing diffeomorphism of $S^1$. On the angular
variable
$\varphi\in [0,2\pi]$, its action is $P:\varphi\mapsto 2\pi-\varphi$.
Classically
(\ref{eq:parity-1}) is a symmetry of the equation of motion
(\ref{eq:cl-e.o.m-1}).

The time-reversal transformation $T$ defined by
\begin{equation}
      T:e^{i\varphi(t)}\mapsto e^{-i\varphi(-t)}
\end{equation}
is also a classical symmetry.

\subsection{Quantum Theory}

In quantum theory, the Hamiltonian $H$ from which one can obtain
(\ref{eq:cl-e.o.m-1}) is
\begin{equation}
      \label{Hamiltonian-1}
      H=-\frac{1}{R} \frac{d^2}{d\varphi^2},
\end{equation}
where the constant $1/R$ has the dimension of energy.

The Hilbert space for a particle on $S^1$ is
\begin{equation}
      \mathcal{H}\equiv L^2(S^1)=\{ \langle \chi,\psi \rangle :=\int_0^{2\pi}
d\varphi
~\bar{\chi} \psi < \infty, \textrm{ for } \chi,\psi\in L^2(S^1)  \}.
\end{equation}
As usual, $\langle \psi,\psi \rangle=\| \psi \|^2$.

Now, the Hamiltonian $H$ has several different domains for which it is
self-adjoint.
They are
labeled by the points $\eta=e^{i\theta}$ of $S^1$. The definition of these
domains
is\footnote{There are also some differentiability (Sobolev) conditions for
$\psi$ in
these
domains.}
\begin{equation}
      \label{eq:domain-sa-1}
      D_\eta=\{ \psi\in\mathcal{H} : \psi(2\pi)=\eta \psi(0)   \}.
\end{equation}

The density matrix $|\psi\rangle\langle\psi|$ associated to
$\psi\in D_\eta$ is a
periodic function of $\varphi$, since $\eta$ cancels out, showing that
(\ref{Hamiltonian-1})
is appropriate for quantum dynamics on $S^1$.

Another way to see that (\ref{eq:domain-sa-1}) is good for quantum dynamics on
$S^1$
is the
following. Let us consider the algebra $\mathbb{C}^\infty(S^1)$ of smooth
functions
on $S^1$.
Then $D_\eta$ is a module for $\mathbb{C}^\infty(S^1)$, that is, if
$f\in\mathbb{C}^\infty(S^1)$ and $\psi\in D_\eta$, then
\begin{equation}
f\psi\in D_\eta.
\end{equation}
As $S^1$ can be recovered from $\mathbb{C}^\infty(S^1)$ as a topological space
by the
Gel'fand-Naimark theorem\footnote{The closure of $\mathbb{C}^\infty(S^1)$ in the
sup-norm gives
a $\mathbb{C}^*$-algebra to which we can apply the Gel'fand-Naimark theorem.}
\cite{gelfand-naimark}, we again see
that (\ref{eq:domain-sa-1}) works out.

All of these remarks go towards solving an old problem of the \emph{Quantum
Baby} described in detail in \cite{bal-quantum-baby}.

\subsubsection{Parity}

Parity $P$ acts on $\psi$ according to
\begin{equation}
      (P\psi)(\varphi)=\psi(2\pi-\varphi).
\end{equation}
Hence, if $\psi\in D_\eta$, then
\begin{equation}
      (P\psi)(2\pi)=\psi(0)=\bar{\eta}\psi(2\pi)=\bar{\eta}~(P\psi)(0),
\end{equation}
or $P\psi\in D_{\bar{\eta}}$, that is,
\begin{equation}
      PD_\eta=D_{\bar{\eta}}.
\end{equation}
The conclusion is that $P$ is anomalous unless $\eta=\bar{\eta}$ or $\eta=\pm
1$. In
terms of $\theta$, the statement is that $P$ is anomalous unless
$\theta=0,\pi\mod
2\pi$.

\subsubsection{Time Reversal}

Since $T$ is an anti-unitary operator,
\begin{equation}
      TD_\eta=D_{\bar{\eta}},
\end{equation}
so $T$ as well is broken, unless again $\eta=\bar{\eta}$ or $\eta=\pm 1$.

Note however that $PT$ preserves $D_\eta$ for all $\eta$,
\begin{equation}
      PTD_\eta=D_\eta.
\end{equation}

Recall that in $1+1$ QED and $3+1$ QCD, the well-known $\theta$-terms also
break $P$ and $T$,
unless $\theta=0,\pi\mod 2\pi$, while $PT$ is always preserved. This coincidence
is
not
accidental. It comes from the fact that $\pi_1(Q)=\mathbb{Z}$ for their
configuration
spaces
$Q$ \cite{bal-book-1,bal-book-fibre}.

\subsubsection{Restoration of $P$ and $T$}

A naive approach to restoration of $P$ and $T$, which however does not work, is
the
following. Consider the case of $P$. For $\psi,\chi\in D_\eta$, we can
declare that the domain of $H$ consists of vectors
of the form $\psi+P\chi$. Since
$\psi$ or $\chi$ can be zero, this means that we would like to declare the
linear
span
$D$ of $D_\eta$ and $PD_\eta$ as the domain of $H$.

This approach does not work as $D$ is not a domain for $H$. An easy way
to
see this
fact is to check that
\begin{equation}
      \langle\psi+P\chi,H(\psi+P\chi) \rangle - \langle H(\psi+P\chi),\psi+P\chi
\rangle
\end{equation}
is not zero for generic $\psi,\chi$. So $H$ is not even symmetric on
$D$.

Another, but different, reason to discard such $D$ is to note that
\begin{equation}
      | \psi+P\chi \rangle \langle \psi+P\chi |
\end{equation}
is not a periodic function of $S^1$ for generic $\psi,\chi$. Thus $D$
is not
adapted
to the quantum particle problem on $S^1$.

Now, if we do not insist that $H$ is always defined, but only the unitary time
evolution $e^{-itH}$ is, then as this is a bounded operator, it is defined on
all of
$\mathcal{H}$, an hence also
on $D$. For this definition of $e^{-itH}$, we can start with $H$ having
domain $D_\eta$, and define $e^{-itH}$ on $D_\eta$
and then extend it to all of $\mathcal{H}$ (see below). However this will not
resolve the second difficulty noted above, as $D$ is still not adapted
to an
underlying $S^1$. Furthermore,
the evolutions $e^{-itH}$ are different if the starting domain is
$D_\eta$ or $D_{\bar{\eta}}$ (if $\eta\neq \bar{\eta}$), for
instance.

Thus such superpositions of vectors to overcome anomalies in $P$ or $T$ do not
work.

There is an alternative though. For $\psi\in D_\eta$, we note that
\begin{equation}
      \Omega=|\psi\rangle\langle\psi|+P|\psi\rangle\langle\psi|P
\end{equation}
has positive trace if $|\psi\rangle$ is not a zero vector, that is,
\begin{equation}
      \Tr \Omega = 2 \langle \psi,\psi\rangle > 0.
\end{equation}
Hence
\begin{align}
      \omega &= \frac{\Omega}{\Tr \Omega}, \nonumber \\
      \Tr \omega & = 1,
\end{align}
is a well-defined state on observables. Moreover it is $P$ and $T$ invariant and
is continuous on $S^1$.

If $K=K^\dagger$ is a (bounded) observable, its mean value in this state is
defined
by
\begin{equation}
      \label{eq:mean-value-1}
      \omega(K)=\Tr K\omega=\frac{1}{\Tr\Omega}\left[ \langle \psi|K|\psi\rangle
+\langle\psi|PKP|\psi\rangle\right].
\end{equation}
Since
\begin{equation}
      \omega(K)=\omega(PKP),
\end{equation}
then $\omega(K)$ is zero for $P$-odd $K$:
\begin{equation}
      \omega(K)=0, ~~~ \textrm{ if } ~ PKP=-K.
\end{equation}

If $P$ were not anomalous, so that $\eta=\pm 1$, then $\psi\in D_\eta$ 
need not be an eigenstate of $P$. So $|\psi\rangle\langle\psi|$ may have no
definite
parity, and
$P$-odd observables $K$ may have non-trivial expectation values
$\langle\psi|K|\psi\rangle$.

As for time-evolution, it is important to keep its group property. So we can
time-evolve $|\psi\rangle$ by $e^{-itH_\eta}$ or $e^{-itH_{\bar{\eta}}}$ to
obtain $|\psi_t\rangle_{\eta}$ or $|\psi_t\rangle_{\bar{\eta}}$. We can then use
(\ref{eq:mean-value-1}) to calculate the mean value of $K$. As this mean value
does depend on $\eta$, we still have two physically distinct choices for
time evolution.

Note that $P$-invariant observables form a subalgebra. 

Our rule
(\ref{eq:mean-value-1}) for expectation values can actually be derived by
restricting $\omega$ to $P$-invariant operators. Thus if $PKP$ is $K$, then
\begin{equation}
      \langle \psi|PKP | \psi\rangle = \langle \psi |K|\psi\rangle =
\frac{1}{2}\left[\langle \psi|PKP | \psi\rangle + \langle \psi |K|\psi\rangle
\right],
\end{equation}
which leads to (\ref{eq:mean-value-1}). We have emphasized the significance of
this result for gauge theories in the introduction. 

All the above remarks are seen to straightforwardly apply to time reversal $T$. 

\subsubsection{Summary}

In the presence of $P$ and $T$ anomalies, we can restore them compatibly with
time
evolution.
We must however work with \emph{impure states} $\omega$ of rank 2. We must work
with
$P$-invariant states and so also $P$-invariant observables.

For anomalous \emph{gauge} symmetries like color, this is actually
good, as it gives the possibility of restoring gauge invariance.

\subsection{What is an Anomaly?}

In the general formulation of quantum theory, it is assumed that any bounded
self-adjoint
operator $K$ is an observable. Being bounded, it is defined on all of
$\mathcal{H}$.
Such $K$
can however mix domains.

Let us consider for example the unitary operator $U_{\eta'}$, with
$\eta'=e^{i\theta'}$,
defined by
\begin{equation}   
\left(U_{\eta'}\psi\right)(\varphi)=e^{i\frac{\theta'}{2\pi}\varphi}
\psi(\varphi).
\end{equation}
Acting with this operator on $D_\eta$, one changes $\eta$ to
$\eta'\eta$,
i.e.,
\begin{equation}
      U_{\eta'} D_\eta = D_{\eta'\eta}.
\end{equation}
Moreover, since $U_{\eta'}$ is a bounded operator, it is defined on all of
$\mathcal{H}$.

Now, the operators
\begin{align}
      K&=\frac{1}{2}\left(U_{\eta'}+ U^\dagger_{\eta'} \right), \\
      K'&= \frac{1}{2i} \left(U_{\eta'}-U^\dagger_{\eta'} \right)
\end{align}
are bounded and self-adjoint. Are they observables?

In fact, the parity operator $P$ is bounded and self-adjoint. Is it an
observable? If
yes, is
its anomaly problem spurious?

A closer examination reveals that in the presence of domain-changing
observables,
there is no
canonical choice for time evolution. Any choice will fail to
commute
with the domain-changing observable. We have already remarked on this point and its relation to spontaneous symmetry breaking. That is so even if it generates a classical
symmetry like $P$.
In the
latter
case, we call the classical symmetry anomalous.

\subsubsection{Extension of $e^{-itH_\eta}$ to all of $\mathcal{H}$}

We begin by solving the eigenvalue problem
\begin{equation}
      \label{Hamiltonian-0}
      H_\eta \psi_n^\eta = E_n \psi_n^\eta.
\end{equation}
The solution is (recalling that $\eta=e^{i\theta}$ and
$\psi^\eta_n\in D_\eta$)
\begin{align}
\label{sol-eigvl-0}
      \psi^\eta_n(\varphi) &=\frac{1}{\sqrt{2\pi}}
e^{i(n+\frac{\theta}{2\pi})\varphi}, \\
      E_n &=\frac{1}{R}(n+\frac{\theta}{2\pi})^2, ~~~ \textrm{with } ~
n\in\mathbb{Z}.
\end{align}

Now, $\{\psi^\eta_n \}$ is a complete set. So any $\chi\in\mathcal{H}$, even if
it
is not in
$D_\eta$, can be expanded in the basis $\{ \psi^\eta_n\}$:
\begin{align}
\label{sol-eigvl-1}
      \chi&=\sum_n a_n \psi_n^\eta \\
      a_n &= \left( \psi_n^\eta,\chi \right).
\end{align}
The expansion converges in norm, that is, 
\begin{equation}
\lim_{N\to\infty} \| \chi - \sum_{|n|\leq N} a_n \psi_n^\eta  \|=0. 
\end{equation}

The time evolution of $\chi$ under $e^{-itH_\eta}$ is
\begin{equation}
\label{time-evol-2}
      \chi_t=e^{-itH_\eta}\chi_0=\sum_{|n|\leq N} a_n e^{-itE_n}\psi_n^\eta,
\end{equation}
for a initial $\chi_0=\chi$. The R.H.S. converges, since $|a_n
e^{-itE_n}|=|a_n|$.

But if $\chi_t\notin D_\eta$, term-by-term differentiation of R.H.S. in
$t$
leads to a
divergent series.

We can illustrate this by considering a periodic $\chi$ and $\eta\neq 1$. Set
\begin{equation}
      \chi(\varphi)=\chi_M(\varphi)=\frac{1}{2\pi}e^{iM\varphi}, ~~~~
M\in\mathbb{Z}.
\end{equation}
Then
\begin{equation}
      \label{modes-1}     
a_n=\frac{1}{2\pi}\int_0^{2\pi}d\varphi~e^{-i(n+\frac{\theta}{2\pi})\varphi}
e^{iM\varphi}=\frac{1}{2\pi} \frac{i}{n+\frac{\theta}{2\pi}-M}\left(
e^{-i\theta}-1
\right).
\end{equation}
With these $a_n$, the series (\ref{sol-eigvl-1}) and (\ref{time-evol-2})
converge
since $|a_n|=O(\frac{1}{n^2})$ as $|n|\to\infty$: 
\begin{equation}
\sum_n |a_n|^2 < \infty.
\end{equation}

But
term-by-term
differentiation of (\ref{sol-eigvl-1}) leads to a divergent series since
$|a_nE_n|=O(|n|)$ as
$n\to\infty$.

The conclusion is that time evolution $U_\eta(t)$ determined by $H_\eta$ is
defined
on all $\mathcal{H}$ (and is continuous in $t$), but is differentiable in $t$
only on
vectors in the domain $D_\eta$ of the Hamiltonian $H_\eta$. If a
classical
symmetry $g$ does not preserve this domain, then $gU_\eta(t)-U_\eta(t)g\neq 0$
on all
of
$\mathcal{H}$, and we say that $g$ is anomalous.

\subsection{Relation to Lagrangian Approach}

In this subsection, we explain how our discussion of anomalies based on domains can be interpreted in conventional terms. The example of the particle on a circle gives a transparent model for this demonstration.

Consider the operator 
\begin{equation}
      U_{\bar{\eta}}: ~~~ \left( U_{\bar{\eta}}\psi\right) (\varphi) = e^{-i\frac{\theta}{2\pi}\varphi} \psi(\varphi). 
\end{equation}
For $\psi\in D_\eta$, then
\begin{equation}
      U_{\bar{\eta}} \psi \in D_1.
\end{equation}
Now, $D_1$ consists of periodic functions and it is invariant under parity. But the new Hamiltonian
\begin{equation}
      H_{\eta}=U_{\bar{\eta}} ~H~U_{\bar{\eta}}^{-1} = \frac{1}{R}\left(-i \frac{\partial}{\partial\varphi}+\frac{\theta}{2\pi}\right)^2
\end{equation}
is not parity invariant.

Using canonical methods, it is easy to show that the Hamiltonian $H_\eta$ comes from a Lagrangian
\begin{equation}
      L_\eta=\frac{R}{2}\dot{\varphi}^2-\frac{\theta}{2\pi}\dot{\varphi}.
\end{equation}

In $L_\eta dt$, $-(\theta/2\pi)d\varphi$ is a topological term. It is closed, but not exact on $S^1$. It is the analogue of the Wess-Zumino-Witten term \cite{bal-book-1} or the topological term in the charge-monopole Lagrangian \cite{unfolding}.

We can also model ``covariant'' and ``consistent'' anomalies of quantum field theory in this model. For this purpose, for clarity, we write $-i(\theta/2\pi)d\varphi$ as a connection:
      \begin{equation}
            A(\varphi) = e^{i\frac{\theta}{2\pi}\varphi} d\left( e^{-i\frac{\theta}{2\pi}\varphi} \right),
      \end{equation}
so that
\begin{equation}
      L_\eta dt = \frac{R}{2}\dot{\varphi}^2 dt - i A(\varphi)
\end{equation}

Note that we can allow any fluctuation in $A$, which is an exact one-form on $S^1$ without affecting the cohomology class of $A$. Such fluctuations will not change the domain $D_\eta$ of the Hamiltonian. Let us allow such fluctuations now.

For that we write
\begin{equation}
      A=-i a(\varphi) d\varphi
\end{equation}
and
\begin{equation}
      \label{Lag-eta}
      L_\eta=\frac{R}{2}\dot{\varphi}^2 - a(\varphi) \dot{\varphi}.
\end{equation}
This Lagrangian defines a model invariant under the ``small'' gauge transformations
\begin{align}
      \label{gauge-transf-1}
      a(\varphi) &\rightarrow a(\varphi) + \frac{\partial \Lambda}{\partial \varphi}, \\
      \Lambda(2\pi) &= \Lambda(0) \mod 2\pi, \label{gauge-transf-2}
\end{align}
as they change (\ref{Lag-eta}) only by a total derivative $-d\Lambda/dt$. Furthermore, it preserves the domain $D_\eta$. Hence they preserve the spectrum of the Hamiltonian. (The meaning of the $\mod 2\pi$ qualification in (\ref{gauge-transf-2}) is that $e^{i\Lambda(\varphi)}$ defines a $U(1)$-valued function on $S^1$.).

If a Maxwell term $F^2(\phi)$ is introduced for $a(\phi)$, the Gauss law reads
\begin{equation}
      \label{Gauss-law-1}
	    \frac{\partial E(\phi)}{\partial\phi} - \frac{\theta}{2\pi} \delta(\phi-\varphi)=0,
\end{equation}
where $E(\phi)$ is the electric field. This is the analogue of the Gauss law in the presence of a point charge at $z(t)$ at time t:
      \begin{equation}
            \frac{\partial E^i(x)}{\partial x^i}+e \delta^3 (x-z(t))=0.
      \end{equation}

The charge $Q$ on $S^1$ is thus given by integrating (\ref{Gauss-law-1}), so that
      \begin{equation}
            Q=E(2\pi)-E(0)=\frac{\theta}{2\pi}.
      \end{equation}
This charge is conserved. But under an anomalous gauge transformation, where the gauge function $\Lambda$ does not fulfill (\ref{gauge-transf-2}), $\theta$ changes. So it is not invariant under such gauge transformations. It is thus the analogue of the ``consistent'' charge. The corresponding ``consistent'' but not gauge invariant current
\begin{equation}
     \frac{\partial E(\phi)}{\partial \phi} - \frac{\theta}{2\pi} \delta(\phi-\varphi) 
\end{equation}
happens to be zero here. The corresponding ``covariant'' gauge invariant current is
      \begin{equation}
            \frac{\partial E(\phi)}{\partial \phi} .
      \end{equation}

\section{Non-abelian Monopoles and Breakdown of Color}

In 't Hooft-Polyakov models, magnetic monopoles are associated with
twisted $G$-bundles on the sphere $S^2_\infty$ at $\infty$. Here
$G$ is the remaining gauge symmetry group after the breaking $G^{(0)}\to G$ by a
Higgs field $\Phi$. This remaining group $G$ is also known as ``global'' or ``large'' gauge group. Furthermore, $S^2_\infty$ refers to a large enough spatial sphere, where $\Phi$
can
be
approximated by its asymptotic value $\Phi_\infty$.

In the unitary gauge, where $\Phi_\infty$ takes a constant value on
$S^2_\infty$, the
$G$-bundle is described by a transition function on a small strip $\theta\in
[\pi/2-\epsilon,\pi/2+\epsilon]$ around the equator of $S^2_\infty$, where
$\theta$
is the
polar angle. This is called a \emph{collar neighborhood} $N_\epsilon$ of the
equator
in $S^2_\infty$.
When $\theta$ lies in $N_\epsilon$ and the azimuthal angle $\varphi$ increases
from
$0$ to
$2\pi$, the transition function $\tau$ maps this curve to a non-contractible
loop in
$G$.

It can happen that the values $\tau(\theta,\varphi)$ taken by $\tau$ are not in
the
center $\mathcal{C}$ of $G$. In that
case $g \tau(\theta,\varphi) g^{-1}\neq \tau(\theta,\varphi)$ for all $g\in G$.
The
group $G$ is then broken.

As examples, consider  $U(2)$ and $U(3)$. The
second group contains the color group $SU(3)$ and the electromagnetic $U(1)$,
since
$U(3)= [SU(3)\times U(1)]/\mathbb{Z}_3$.

Let us first consider $U(2)=\left( SU(2)\times U(1) \right)/\mathbb{Z}_2$. We
work in its
two-dimensional (faithful) representation by unitary matrices. Then the choice
\begin{equation}
\tau(\theta,\varphi)=e^{\frac{i}{2}\sigma_3 \varphi }e^{\frac{i}{2}\varphi},
\end{equation}
where $\sigma_3$ is the third Pauli matrix, gives a non-contractible loop
in $U(2)$, which is not entirely contained in its center $U(1)$. The homotopy
class
of this loop generates $\pi_1[U(2)]=\mathbb{Z}$.

A similar discussion applies to $U(3)=\left[SU(3)\times U(1)
\right]/\mathbb{Z}_3$.
In its three-dimensional irreducible representation, the diagonal matrix
$Y=\frac{1}{3}(1,1,-2)$ is in the Lie algebra $u(3)$ of
$U(3)$. The transition function $\tau$ defined by
\begin{equation}
      \tau(\theta,\varphi)=e^{iY\varphi}e^{-i\frac{2\pi}{3}\varphi}
\end{equation}
is a non-contractible loop which is not contained in the center of $U(3)$. So,
for a
generic $g\in U(3)$,
\begin{equation}
      g \tau(\theta,\varphi) g^{-1}\neq \tau(\theta,\varphi)
\end{equation}
in the entire collar neighborhood around the equator. Thus, global $SU(3)$ color cannot
be
implemented.

In \cite{bal-book-1,bal-book-fibre}, it was shown that each such $\tau$ characterizes a domain
$D_\tau$ of say
the Dirac Hamiltonian $H^{D}$. Moreover, global $SU(3)$ color becomes anomalous because
its
action
changes $D_\tau$ to $D_{g\tau g^{-1}}$.

We can now restore color as a symmetry by following the procedure described in
the
last
section. Let $|\chi\rangle_\tau$ be a state vector for the transition function
$\tau$.
This
defines its gauge. It is in the domain $D_\tau$.

Suppose a $g\in G$, it acts on $\tau$ by conjugation
\begin{equation}
      \left( g\tau g^{-1} \right) (\theta,\varphi) = g \tau(\theta,\varphi)
g^{-1}.
\end{equation}
So
\begin{equation}
      gD_\tau = D_{g\tau g^{-1}}.
\end{equation}

Following section 2, we thus consider
\begin{equation}
\Omega=\int_G d\mu(g)~g |\chi\rangle_\tau~_\tau\langle \chi | g^\dagger = \int_G
d\mu(g)~
|\chi\rangle_{g\tau g^{-1}}~_{g \tau g^{-1}}\langle \chi |,
\end{equation}
where $d\mu(g)$ is the Haar measure on $G$. 

This $\Omega$ is a positive $G$-invariant
operator, so that
\begin{equation}
      \omega=\frac{\Omega}{\Tr \Omega}
\end{equation}
is a $G$-invariant state.

Let $H_\tau$ be the Hamiltonian with domain $D_\tau$. On the
intersection
\begin{equation}
      \bigcap_{g\tau g^{-1},~g\in G} D_{g\tau g^{-1}} = D^0
\end{equation}
of these domains, the Hamiltonian $H_{g\tau g^{-1}}$ coincide
\begin{equation}
      H_{g\tau g^{-1}}|_{D^0}=H_\tau,
\end{equation}
for all $g\in G$. Also,
\begin{equation}
      g e^{-it H_\tau} g^{-1}=e^{-i t H_{g \tau g^{-1}}}.
\end{equation}

We now define $\Omega_t$ at time $t$ by
\begin{equation}
      \Omega_t = \int_G d\mu(g) e^{-it H_{g \tau g^{-1}}} ~ |\chi\rangle_{g\tau
g^{-1}}~_{g
\tau g^{-1}}\langle \chi |~e^{it H_{g \tau g^{-1}}},
\end{equation}
with $\Omega_0$ being $\Omega$. Now, $\Omega_t$ is positive and $G$-invariant.
It
gives the $G$-invariant state
\begin{equation}
      \label{time-evol-dens-5}
      \omega_t=\frac{\Omega_t}{\Tr \Omega_t}.
\end{equation}
The state $\omega_t$ is impure.

\subsection{Is Color Confinement a Domain Problem?}

Suppose that there is no twisted $SU(3)$- or more generally twisted $G$-bundle
on
spatial slices, so that state vectors $|\chi\rangle$, which are color ($G$-)
non-singlets in the
domain of the Hamiltonian. Suppose though that there is ``confinement'' in the
sense
that we
observe only $SU(3)$-invariant operators $K$. Such (bounded) operators  form an
algebra
$\mathcal{A}$. Then $|\chi\rangle\langle \chi|$ (with $\langle
\chi|\chi\rangle=1$)
restricted
to $\mathcal{A}$ is in fact an impure state like the one we discussed before.
That is because we can trace over $|\psi\rangle\langle\psi |$ the color degrees
of freedom. This point was emphasized by Akant et al \cite{rajeev}.

To see this explicitly, let $U(g)$ be the unitary operator implementing $G$.
Then for
$K\in\mathcal{A}$,
\begin{align}
\langle \chi |K|\chi\rangle &=\frac{1}{V}\int_G d\mu(g) \langle
\chi|U(g)^\dagger K
U(g) |\chi\rangle, \\
V &=\int_G d\mu(g),
\end{align}
or
\begin{align}
      \Tr K|\chi\rangle\langle \chi| &= \Tr K\omega, \\
      \omega &=\frac{1}{V} d\mu(g) U(g) |\chi\rangle\langle \chi| U(g)^\dagger.
\end{align}

Since the Hamiltonian $H$ must be a $G$-singlet if $H$ is to display
confinement, we
can
evolve $\omega$ for time $t$ in a conventional way,
\begin{equation}
      \label{time-evol-dens-7}
      \omega_t=e^{-itH}\omega_0 e^{it H},
\end{equation}
with $\omega_0=\omega$. The previous formula (\ref{time-evol-dens-5}) reduces to
(\ref{time-evol-dens-7}) when there is no domain problem.

However, we were led to the singlet states $\omega_t$ of
(\ref{time-evol-dens-5})
because of
domain problems caused by non-abelian monopoles. Is this a first step towards a
proof of confinement?

Discussions of confinement also speculate that colored states have infinite mean
energy. That is also the case here if this conjecture is suitably interpreted.
Thus,
first consider $e^{-itH_\tau}$, $H_\tau$ being the Hamiltonian with domain
$D_\tau$. It can be defined on all $\mathcal{H}$ including vectors
$|\chi\rangle_{g\tau
g^{-1}}$, with $g\tau g^{-1}\in D_{g\tau g^{-1}}\neq D_\tau$.
But
\begin{equation}
i\frac{d}{dt} _{g\tau g^{-1}}\langle \chi| e^{-itH_\tau} |\chi\rangle_{g\tau
g^{-1}}|_{t=0}
\end{equation}
diverges.

We can show this by the parity example of section 2, but the result seems to be
generic. Thus from (\ref{sol-eigvl-1}),(\ref{Hamiltonian-0}),
(\ref{sol-eigvl-0}) and
(\ref{modes-1}), and also
\begin{equation}
      \label{chiM-ener-av-0}
      \langle \chi_M | e^{-itH_\eta}|\chi_M\rangle = \sum_{n} |a_n|^2
e^{-itE_n},
\end{equation}
it follows that
\begin{align}
      \label{chiM-ener-av-1}
      E_n &= \frac{1}{R} (n+\frac{\theta}{2\pi})^2 \\
      a_n &= \frac{1}{2\pi} \frac{1}{n+\frac{\theta}{2\pi}-M} \left(
e^{-i\theta}-1
\right)
\label{chiM-ener-av-2},
\end{align}
showing that (\ref{chiM-ener-av-0}) is not differentiable in $t$ or that the
mean
energy
$\langle\chi_M|H_\eta|\chi_M\rangle$ is infinite.

This is perhaps a mechanism which contributes to confinement. But for further
progress, we still
need non-abelian colored monopoles associated with reasonable length scales.
Unfortunately, we
know of none. GUT monopoles seem too small for our purpose. If
the length
scale of quark confinement is $10^{28}\textrm{cm}^{-1}$, then it is hard to
understand the low
energy success of the quark model.

\section{On the Genericity of Gauge Anomalies}

Let $\hat{G}$ be a gauge group for a quantum system based on a Hamiltonian $H$.
\emph{By
definition}, all observables, including $H$, commute classically with $\hat{G}$.

In quantum theory, typically, the identity component $\hat{G}_0$ of $\hat{G}$ is
required to act
trivially on quantum states by virtue of a Gauss law. The group
$\hat{G}/\hat{G}_0=G$
can then
act by an unitary irreducible representation (UIRR) $\rho$ on the quantum
states.

As an example, consider QCD. There, for $\hat{G}$, we can consider
$\mathcal{G}^\infty(SU(3))$,
the group of maps from $\mathbb{R}^3$ to $SU(3)$, which reduce to identity at
spatial
infinity. Its identity component $\mathcal{G}^\infty_0(SU(3))$, being generated
by
Gauss law,
acts trivially on quantum states. Now,
$\mathcal{G}^\infty(SU(3))/\mathcal{G}^\infty_0(SU(3))=\pi_3(SU(3))=\mathbb{Z}$.
It
has UIRR's
$\rho\equiv \rho_\theta$ with $\rho_\theta(n)=e^{in\theta}$ for
$n\in\mathbb{Z}$. The
angle
$\theta$ is fixed in a given QCD theory. 

In quantum gravity based on asymptotically flat space-times, the approach of
diffeomorphisms
$D^\infty(M)$ of the spatial slice $M$ which become asymptotically identity
plays a
role
similar to $\mathcal{G}^\infty(SU(3))$. Its identity component $D^\infty_0(M)$
acts
trivially
on quantum states, while the discrete group $D^\infty(M)/D^\infty_0(M)$ acts by
some
UIRR $\rho$
on quantum states.

There are examples of a different sort from molecular physics \cite{bal-book-1}.
In the
Born-Oppenheimer
approximation, the family of nuclear orientations which serves as the
configuration space
 $Q$ for
rotational excitations is $SU(2)/G$, where $G$ is a subgroup of $SU(2)$.
It may
be
discrete giving rise to a Platonic solid \cite{bal-simoni-witt}, $U(1)$ or
$\mathbb{Z}_4 \ltimes U(1)$. If
$U(1)=\{e^{i\theta
\sigma_3/2},
0\leq\theta \leq 4\pi \}$ and $\mathbb{Z}_4=\{z=i\sigma_2: z^4=e \}$, then it is
generated by $\langle
e^{i\theta \sigma_3/2},i\sigma_2 \rangle$.

In time-reversal invariant systems, if the value $\mathbf{k}_0$ of momentum
$\mathbf{k}$ is
time-reversal invariant, then the sphere $\{\mathbf{k}:
|\mathbf{k}-\mathbf{k}_0|^2=1
\}$ can
support a $\mathbb{Z}_2$-bundle
\cite{mukunda-1,mukunda-2,mukunda-3,bal-amilcar}. The $\mathbb{Z}_2$ is
generated
by the
square of
the time-reversal transformation $T$. According to Wigner \cite{wigner}, $T^2$ is
either $+\mathbb{1}$ or $-\mathbb{1}$. $T$ can act on quantum states by either
of these
two UIRR's.
Since observables necessarily commute with the square of time-reversal
transformation,
$\mathbb{Z}_2$ is a gauge group. These bundles occur in discussions of
topological
insulators
\cite{kane-hasan}.

Thus there are plenty of gauge groups $G$ and many are non-abelian.

Let us call the effective gauge group after possible Gauss-law constraints are
accounted for as
$G$. As explained above, it is the group which can act by non-trivial
representations
$\rho$ on
quantum states.

Now if $\rho(g)$ is the unitary operator representing $g\in G$ on quantum
states,
then $\rho$
also gives a representation of the entire group algebra $\mathbb{C}G$ of $G$. If
$\sum_g c(g) g
\in \mathbb{C}G$, where $c(g)\in \mathbb{C}$, then its operator is $\sum_g c(g)
\rho(g)$. This
representation incidentally is a $*$-representation:
\begin{equation}
      *: \sum_g c(g) g \to \sum_g \overline{c(g)} g^{-1}
\end{equation}
on $\mathbb{C}G$ goes over to the adjoint operations in the representation
\begin{equation}
      \rho\left( \sum_g \overline{c(g)} g^{-1}\right)=\left(\sum_g c(g) \rho(g)
\right)^\dagger,
\end{equation}
since $\rho(g)^\dagger=\rho(g^{-1})$.

Now \emph{all observables must commute with $\mathbb{C}\hat{G}$, the gauge group
algebra of
$\hat{G}$, and in particular with $\mathbb{C}G$.} That is the meaning of gauge
invariance.
\emph{But if $G$ and hence $\mathbb{C}G$ are non-abelian, only the center
$\mathcal{C}(\mathbb{C}G)$ of $\mathbb{C}G$ commutes with every element of
$\mathbb{C}G$. If
$G$ is abelian, the $\mathcal{C}(\mathbb{C}G)=\mathbb{C}G$, but that is not the
case if $G$ is non-abelian.}

Thus if $G$ is a finite group, its center has the basis \cite{bal-book-groups}
\begin{equation}
      e_\alpha=\sum_g \chi_\alpha(g) g,
\end{equation}
where $\chi_\alpha$ is the character in the irreducible representation
$\rho_\alpha$.
If instead $G$ is a
compact Lie group, its center is spanned by the Casimir invariants. In either of
these cases of
interest, $\mathcal{C}(\mathbb{C}G)$ is an \emph{abelian} algebra.

Since $\mathcal{C}(\mathbb{C}G)$ lies in the center of the entire algebra of
observables, in a
given representation of the latter, elements of $\mathcal{C}(\mathbb{C}G)$ have
a
fixed value.
Fixing $e_\alpha$ means fixing the irreducible representation\footnote{The
$e_\alpha$'s after a
normalization become orthogonal projectors.} while for Lie groups $G$, we will
be
fixing its
Casimirs.

Thus general considerations fix \emph{only} the UIRR $\rho$ of $\mathbb{C}G$.
The
$\rho(g)$ acts
on a Hilbert space $\mathcal{H}$ by a unitary representation, so we can choose
a
complete set
spanning $\mathcal{H}$ in the form
\begin{equation}
      |\sigma\rangle \otimes |\psi\rangle \equiv |\sigma,\psi\rangle,
\end{equation}
where
\begin{equation}
      \label{eq:uir-2}
      \rho(g)|\sigma,\psi\rangle=|\sigma',\psi\rangle \rho(g)_{\sigma' \sigma},
\end{equation}
on denoting the matrix of $\rho(g)$ by the same symbol.

Now, elements of $\rho(\mathcal{C}(\mathbb{C}G))$ have exactly the same value on
$|\sigma,\psi\rangle$, for every $\sigma\in\mathcal{C}(\mathbb{C}G)$, with
$\rho$
being
irreducible. So $\mathcal{C}(\mathbb{C}G)$ does not mix different values of
$\sigma$,
nor does
any other observable as it commutes with $\rho(G)$. So we have to ``gauge fix"
the
redundancy
in the multiplicity of $\sigma$ if possible. 

We are assuming that the dimension of
$\rho(G)$
is larger than one, otherwise $\rho(\mathbb{C}G)$ is abelian.

One possibility that may occur is that we can fix the value for $\sigma$, and
choose
a domain
for observables in the span of $\{|\psi\rangle \}$. This may be possible with
observables
acting just on $|\psi\rangle$. The $\psi$'s  are typically functions on a
classical
configuration space $Q$, so that in this case the quantum vector bundle over $Q$
is
trivial.
Physical predictions in this case do not depend on $\sigma$.

Instead of working with vector states, we can also work with density matrices
\begin{equation}
      \label{density-matrix-4}
      \sum_\sigma \frac{|\sigma,\psi\rangle\langle \sigma,\psi|}{\Tr
|\sigma,\psi\rangle\langle
\sigma,\psi|}.
\end{equation}
Such states are more like our construction in section 2 and treat all $\sigma$
democratically.
However, on observables, both approaches are equivalent when the bundle is
trivial.

Note also that $G$ acts on (\ref{density-matrix-4}) by the identity
representation\footnote{The co-unit for its Hopf algebra
\cite{bal-book-groups}.},
while if we
gauge fix $\sigma$, the $G$-action changes the gauge, but harmlessly.

When the bundle is twisted, we cannot proceed in this manner. In that case, we
cover
$Q$ by
contractible open sets $Q_\alpha$,
\begin{equation}
      Q=\bigcup_\alpha Q_\alpha.
\end{equation}
In each $Q_\alpha$, we choose a section 
\begin{equation}
      \label{eq:section-0}
      \sum_\sigma \chi_\sigma^{(\alpha)} |\sigma,\psi\rangle,
\end{equation}
where $\chi_\sigma^{(\alpha)}$ are smooth functions on $Q_\alpha$. In the
overlap
$Q_{\alpha\beta}=Q_\alpha\cap Q_\beta$, we have a transition function
$U_{\alpha\beta}$, which at $q\in Q_{\alpha\beta}$ gives an element $\rho(g)$,
$g\in
G$, 
\begin{equation}
      U_{\alpha\beta}\in \rho(G), ~~~ q\in Q_{\alpha\beta},
\end{equation}
in a self-evident notation. Then the vectors (\ref{eq:section-0}) and
\begin{equation}
      \sum_\sigma \chi_\sigma^{(\beta)} |\sigma,\psi\rangle
\end{equation}
are related by $U_{\alpha\beta}$ over $Q_{\alpha\beta}$: 
\begin{equation}
      \label{eq:trans-fn-1}
      \sum_\sigma \chi_\sigma^{(\alpha)} |\sigma,\psi\rangle =
U_{\alpha\beta}\sum_\sigma
\chi_{\sigma}^{(\beta)}|\sigma,\psi\rangle ~~~ \textrm{ on } Q_{\alpha\beta}.
\end{equation}
There are also consistency conditions on $U_{\alpha\beta}$ which lead to
\v{C}ech
cohomology
\cite{Nakahara,Steenrod}.

If there exist $U_\alpha$'s which are $\rho(G)$-valued smooth functions on
$Q_\alpha$
such
that 
\begin{equation}
      \label{eq:trans-fn-2}
      U_{\alpha\beta}=U_\alpha^{-1}U_\beta ~~~ \textrm{ on } Q_{\alpha\beta}, 
\end{equation}
then we can reduce $U_{\alpha\beta}$ to the constant function on
$Q_{\alpha\beta}$
with value
${\mathbb 1}$ by choosing different sections, namely
\begin{equation}
      U_\alpha\sum_{\sigma} \chi_\sigma^{(\alpha)} |\sigma,\psi\rangle ~~~
\textrm{
on }
Q_{\alpha}.
\end{equation}
But such $U_\alpha$ may not exist. In that case, the vector bundle is said to be
``twisted".

The choice of sections on $Q_\alpha$ is a ``gauge choice". It also goes towards
fixing the
domain of the Hamiltonian.

If the vector bundle is twisted, we cannot say that the action of $\rho(g)$
preserves
the
transitions functions. As the domain of the Hamiltonian is determined precisely
by
these
transition functions, we cannot say that $\rho(g)$ preserves the domain. If it
does not,
we say that $G$
is anomalous \cite{esteve-1, esteve-2}.

More generally, there can be a classical symmetry like parity $P$ which is not
part of $\rho(\mathbb{C}G)$. If it does not preserve the domain, that is, the
transition
functions,
then this symmetry is anomalous.

If $G$ is non-abelian, only the elements of $G$ commuting with all
$U_{\alpha\beta}$
preserve
the domain. The rest are anomalous.

In QCD, the global symmetry group $SU(3)$ can be regarded as the group of
constant
maps from
$\mathbb{R}^3$ to $SU(3)$. Since $SU(3)\cap \mathcal{G}^\infty=\{e\}$, they are
not
``gauge
transformations" as per the considerations hitherto. We should really enlarge
$\mathcal{G}^\infty$ to $\mathcal{G}$, which are smooth maps from $\mathbb{R}^3$
to $SU(3)$
which
approach a constant value in $SU(3)$ at infinity (that is, when $|\vec{x}| \to
\infty$). In
that case
$SU(3)$ is part of the gauge group. What we have proved in
\cite{mukunda-1,mukunda-2,mukunda-3} is
that its
action changes
the transition functions and hence the domain of the Hamiltonian in the presence
of
non-abelian
monopoles. Hence $SU(3)$ of color is anomalous in the presence of these
monopoles.

We conclude this section by listing examples where twisted bundles with
non-abelian
gauge
groups occur. A proper investigation of the physics and mathematics of these
bundles
from a
physical perspective does not exist.

\subsection{Examples}

\subsubsection{From Molecular Physics}

As mentioned above, the rotational degrees of freedom of a molecule are
described by
the
configuration space $Q=SU(2)/G$, where $G$ is a subgroup of $SU(2)$
\cite{bal-simoni-witt,bal-book-1}. Since
$SU(2)\neq Q\times G$, the principle bundles $G\to SU(2)\to SU(2)/G$ are
\emph{all} twisted when $G\neq \{ e\}$. There are plenty of molecules with
$\rho(G)$ non-abelian.

We will illustrate our general considerations from such $Q$ in the next section.

\subsubsection{Parastatistics, Braid Group}

The configuration space $Q$ of $N$ identical particles on $\mathbb{R}^d$ is
\begin{equation}
      Q=\{[q_1,...,q_N]: q_i\in\mathbb{R}^d, q_i\neq q_j, ~~ \textrm{ if } i\neq
j\},
\end{equation}
where $[q_1,...,q_N]$ is an \emph{unordered} set \cite{bal-book-1,
Leinaas,birman-book}:
\begin{align}
      \label{unordered-para-1}
      [q_1,q_2,...,q_N]&=[q_{s(1)},q_{s(2)},...,q_{s(N)}] \\
      s &\in S_N, \nonumber
\end{align}
$S_N$ being the permutation group of $N$ particles. It is
(\ref{unordered-para-1}) which enforces the particle identity. Thus $Q$ consists
of $N$ points of $\mathbb{R}^d$ of cardinality $N$.

In quantum theory, for $d\geq 3$, the group $S_N$ arises as the ``gauge" group
commuting with all observables. If $\rho(S_N)$ is abelian, which is the case
\emph{only}  for bosons and fermions, there is no problem in implementing
it on vector states. But if $\rho(S_N)$ is non-abelian, gauge fixing in order
to eliminate the redundant vectors in the representation space leads to anomalies.

For $d=2$, $S_N$ is replaced by the braid group $B_N$
\cite{birman-book,bal-book-1}, allowing
the possibility of fractional statistics. Its non-abelian representations have
recently occurred in discussions of quantum Hall effect at the filling fraction
$\nu=5/2$ \cite{willett}, topological quantum computing \cite{nayak} and the
Kitaev model \cite{kitaev}. If $\rho(B_N)$ is non-abelian, it cannot act on
properly gauge fixed
quantum states.

\subsubsection{Non-abelian Monopoles Break Color}

We have already discussed this issue in section 3 above.

\subsubsection{Mapping Class Groups of Geons}

The mapping class groups here are the groups $D^\infty/D^\infty_0$ already
defined
above for
the Friedmann-Sorkin spatial slices supporting topological geons. They are
discrete,
but are
non-abelian for appropriate slices
\cite{friedman-witt,friedman-sorkin-1,friedman-sorkin-2}. In these cases,
if
$\rho(D^\infty/D^\infty_0)$ is non-abelian, there might appear quantum diffeo
anomalies. We discuss this issue elsewhere \cite{bal-amilcar}.

\section{On Molecular Configuration Spaces}

We will adapt the discussion of \cite{bal-simoni-witt} regarding quantum
theories on
$Q=SU(2)/G$, 
with $G$ a subgroup of $SU(2)$ for illustrating our preceding remarks.

Quantization on $Q$ can conveniently start from its universal cover $SU(2)$ and
functions on
$SU(2)$. The latter are spanned by the components of rotation matrices
$D^j_{\lambda
\mu}$,
with $j\in\mathbb{Z}^+/2$, $\lambda,\mu\in [-j,-j+1,...,j]$, where the scalar
product
is
\begin{align}
      \langle D^{j'}_{\lambda'\mu'},D^j_{\lambda\mu} \rangle &=\int_{s\in SU(2)}
d\mu(s)~\bar{D}^{j'}_{
\lambda'\mu'}(s)
D^j_{\lambda\mu}(s),
\end{align}
where $d\mu(s)$ is the
invariant $SU(2)$ measure (with volume of $SU(2)$ equal to $16\pi^2$, say).
With this scalar product, this space of
functions on
$SU(2)$ generates a Hilbert space.

On functions $f$ on $SU(2)$, there is a left- and a right-action $U_{L,R}$ of
$SU(2)$
defined
by
\begin{align}
      \left( U_L(t) f \right)(s) &= f(t^{-1} s), \\
      \left( U_R(t) f \right)(s) &= f(st), \\
      s,t &\in SU(2) \nonumber.
\end{align}
These actions commute: 
\begin{equation}
      U_L(s)U_R(t)=U_R(t)U_L(s).
\end{equation}

The gauge group $G$ and its group algebra $\mathbb{C}G$ act on the \emph{right},
that is, by the
representation $U_R$. The observables lie in $\mathbb{C}U_L(G)$, so that they
commute with the
gauge transformations $U_R(G)$ and its group algebra $\mathbb{C}U_R(G)$.

We take $U_R$ to be a UIRR. Now,
\begin{equation}
      D^j_{\lambda\mu}(st)=D^j_{\lambda\mu'}(s)D^j_{\mu'\mu}(t),
\end{equation}
so that to obtain an irreducible action of $G$, we must restrict the second
index
to a
suitable
subset.

For example if $G=\mathbb{Z}_N=\{e^{i\frac{2\pi}{N}m\sigma_3}: m=0,1,...,N-1
\}$,
then
\begin{equation}
\label{eq:section-3}
D^j_{\lambda\mu}(se^{i\frac{2\pi}{N}m\sigma_3})=D^j_{\lambda\mu}(s)e^{i\frac{
4\pi}{N}
m\mu}
\end{equation}
remembering that $\mu$ is associated with eigenvalues for $\sigma_3/2$. So for
$\mu\pm 1/2$,
\begin{equation}
      e^{i\frac{2\pi}{N}\sigma_3} \to e^{\pm i \frac{2\pi}{N}}.
\end{equation}
These two representations may or may not be equivalent depending on $N$.

For general $\mu$ the representations are
\begin{equation}
      e^{i\frac{2\pi}{N}\sigma_3} \to e^{i \frac{4\pi}{N}\mu}.
\end{equation}
So
\begin{equation}
	    \mu=\frac{1}{2}+\frac{N}{2}k, ~~~~ k\in\mathbb{Z}
\end{equation}
also give the representation
\begin{equation}
      e^{i\frac{2\pi}{N}\sigma_3} \to e^{+ i \frac{2\pi}{N}}.
\end{equation}
For this UIRR, then, the wave functions are spanned by
\begin{equation}
      \{ D^j_{\lambda,\frac{1}{2}+\frac{N}{2}k}: k\in\mathbb{Z} \}.
\end{equation}

For specificity, we focus on the UIRR $e^{i 2\pi/N~\sigma_3}\to e^{i 2\pi/N}$.
Using (\ref{eq:section-3}), we see that a subset of $\mu$'s, call it $\{\nu\}$,
carry this
UIRR. Then the space spanned by $\{D^j_{\lambda \rho}: \rho\in\{\nu\} \}$ is
invariant under observables. We can reduce this further and fix $\rho$ to a
particular value $\rho_0\in\{\nu\}$ or if one prefers, consider the span of
$\sum c_\rho D^j_{\lambda \rho}$ for fixed $c_\rho\in\mathbb{C}$.

To present this basis in terms of transition functions, we must cover $SU(2)/G$
by
contractible
open sets $Q_\alpha$. Then on $Q_\alpha$, there is a global section. That is,
for
$q\in
Q_\alpha$, we can pick an element $s_\alpha(q)\in SU(2)$ ``in the fiber over"
$q$ smoothly.
More generally, we
can choose a section $s_\alpha(q)g_\alpha(q)\in SU(2)$, with $g_\alpha(q)\in G$.

Now suppose that we choose to work with the span of $D^j_{\lambda
\rho_0}(s_\alpha(q)g_\alpha(q))$ over $Q_\alpha$. Then the sections over
$Q_\alpha$
are
\begin{equation}
      D^j_{\lambda \rho_0}(s_\alpha(q))U_R(g_\alpha(q)),
\end{equation}
where $U_R(g_\alpha(q))$ is a phase. 

The first factor here corresponds to
$|\psi\rangle$ in (\ref{eq:uir-2}), the second to the factor with $\sigma$.

Now consider $U_{\alpha\beta}$. In $U_{\alpha\beta}$, $s_\alpha(q)$ and
$s_\beta(q)$
can differ only by the action of the group, so that
\begin{equation}
      s_\alpha(q)=s_\beta(q) g_{\beta \alpha}(q),
\end{equation}
with $q\in Q_\alpha$ and $g_{\beta\alpha}(q)\in G$. Hence
\begin{equation}
      D^j_{\lambda \rho_0}(s_\alpha(q))U_R(g_\alpha(q))=D^j_{\lambda
\rho_0}(s_\beta(q))U_R(g_\beta(q))U_R(g_{\beta\alpha}(q)).
\end{equation}
The last factor $U_R(g_{\beta\alpha}(q))$ regarded as the evaluation at $q$ of a
function with values in $U_R(G)$ gives the $U_{\alpha \beta}$ of
(\ref{eq:trans-fn-1}).

In the abelian example, there is no problem of implementing $U_R(g)$ for any
$g\in
G$, as they preserve the transition functions. Indeed as $G$ is abelian, $G\in
\mathcal{C}(\mathbb{C}G)$.

But there can still be classical symmetries which can change $U_{\alpha \beta}$.
In
particular, parity $P$ and time-reversal $T$ can do so. In
\cite{bal-simoni-witt}, it was
shown
that
$P$ and $T$ are \emph{not} violated if and only if
\begin{equation}
      \label{parity-time-cond}
      U_R(e^{i \frac{4\pi}{N}\sigma_3})=\pm {\mathbb 1}.
\end{equation}
Otherwise they are violated.

The group $\mathbb{Z}_N$ occurs as $G$ (called $H^*$ in \cite{bal-simoni-witt})
for
pyramidal
molecules. There are pyramidal molecules where (\ref{parity-time-cond}) is not
fulfilled. Their quantum theories violate $P$ and $T$. But just like QCD, $PT$
is not
anomalous in quantum theories.

The groups $D^*_{4N}$, with $N\in\mathbb{Z}$, is the gauge group $G$ for
``staggered"
and ``eclipsed" configurations such as those of ethane \cite{bal-simoni-witt}. 

The group $D^*_8$
has the following elements:
\begin{equation}
      D^*_8=\{ \pm{\mathbb 1},\pm i\tau_i\}\subset SU(2).
\end{equation}
It is the ``symmetry group" or the gauge group leaving the shape of the biaxial
nematic invariant.

Reference \cite{bal-simoni-witt} shows that molecules with $N$ even do not
violate $P$ or $T$.

But
$D^*_{4N}$ are all non-abelian for $N\geq 2$. If $D^*_{4N}$ has $K$ UIRR's, then
the
center $\mathcal{C}(\mathbb{C}D^*_{4N})$ is of dimension $K$. For a generic UIRR
$U_R$, only $U_R(e_\alpha)$, $e_\alpha\in\mathcal{C}(\mathbb{C}D^*_{4N})$ and
their linear combinations are well-defined in a
quantum theory, and we cannot implement the UIRR's $U_R$ of $D^*_{4N}$.

\section{Acknowledgement}

APB and ARQ would like to thank Prof. Alvaro Ferraz (IIP-UFRN-Brazil) for the
hospitality at IIP-Natal-Brazil, where part of this work was carried out. APB also acknowledges CAPES for the financial support during his stay at IIP-Natal-Brazil. APB is
supported by DOE under grant number DE-FG02-85ER40231. ARQ is supported by CNPq under process number 307760/2009-0.

\end{document}